\documentstyle[amsfonts,aps]{revtex}

\def\bigone{\hbox{1 \hskip -5.5pt  I}}

\begin{document}
%\twocolumn 
\draft   

\title{Coherent quasielastic neutron scattering and correlations
between rotational jumps of molecules on a periodic lattice} 
\author{Gerrit Coddens}

\address{Laboratoire des Solides Irradi\'es, Ecole Polytechnique,\\
F-91128-Palaiseau CEDEX, France}   % Declares the author's name.
\date{3rd December 2002}   % Deleting this command produces today's date.

\maketitle
%\widetext 
\begin{abstract}
We previously (G. Coddens, Phys. Rev. {\bf 63}, 064105 (2001)) 
derived a theorem about the {\em coherent} 
quasielastic neutron-scattering
signal from a $d$-dimensional lattice of $N$  molecules that are undergoing 
rotational jump diffusion (around an $n$-fold axis), 
assuming that there are no correlations between the molecules.
In the present paper molecular correlations are treated, but only
in the sense that several molecules could reorient
simultaneously as in a cog-wheel mechanism.
Moreover, we do not examine the possibility that
the relaxation times of
these combined reorientations
could depend on details of the local environment created by 
the neighbouring molecules.
Finally also an ergodicity condition has to be fulfilled.
Admitting for all these assumptions
we can show that the correlations do not affect the coherent 
quasielastic scattering pattern in the following sense:
The functions of $Q$ that 
intervene in the description
of the intensities remain unaltered,
while the functions of $\omega$ can undergo a 
renormalization of the time scales.
The latter changes cannot be detected
as the time scales that would occur
if the dynamics were independent
are not available for comparison.
In other words: Coherent quasielastic neutron scattering
is not able to betray the existence of 
correlations of the restricted type that occur in our model.
The assumptions that underly the model we present were made to
allow a mathematically rigorous calculation of the scattering function.
Other, perhaps more realistic cases may entail correlations 
of a type that is too difficult to be treated
rigorously with our method of calculation.
But our result presents an important non-trivial counterexample
 to show that the absence 
of a clue for the presence
of correlations in the data is not a sufficient criterium to
conclude that such correlations are indeed absent.
\end{abstract}

%\narrowtext

\section{Introduction}

In a recent paper\cite{PRB} we calculated the coherent quasielastic 
neutron-scattering signal\cite{Bee,Springer,Hempelmann}
that results from $n$-fold rotational jumps of $N$ 
molecules periodically arranged on
the sites of  a
$d$-dimensional  solid-state lattice, by applying a previously 
described method.\cite{JdP} 
{\em The jumping molecules
themselves do not need to have $n$-fold symmetry}. 
As a physical visualization of the present problem we could mention the 
array of nonadecane molecules hosted by an urea 
inclusion compound.\cite{Toudic}
(Such molecules have CH$_{3}$ end groups of 
three-fold symmetry, but the entire molecule
has no such symmetry).

In a first approach, we assumed that the 
molecules are jumping independently, 
leaving open the question: What could 
happen if the relaxational motions of
neighbouring molecules cease
to be uncorrelated?
We anticipated that in that case coherent quasielastic neutron 
scattering might be able
to evidence possible correlations.
In the present paper we address this possibility, 
understanding the word correlation in the sense that within 
connected sets of a given size and shape all molecules  
jump simultaneously in some phase or anti-phase relations, 
like in a cog-wheel mechanism.
In the paper we will use the word ``clusters'' to refer to these sets.
Assuming further that
the relaxation times for these combined jumps do not 
depend on the momentaneous orientations
of the molecules  inside (and outside) the cluster,
we come to the astounding (and disappointing) conclusion that such
correlations do not leave a fingerprint
in the structure factors of the ensuing coherent dynamical signal.
(See the abstract or below for a more precise formulation
of the restrictions).
In other words: It is impossible to extract  information on
the presence and/or the nature of 
such correlations from  coherent quasielastic neutron
scattering signals alone.

Let us point out where the surprise lies. 
Quite often, if we calculate the total structure factor 
for such dynamics within a
single isolated cluster,  we will be able to detect 
the presence of such correlations, as
 we do not 
find the same answer as for the jump dynamics
of a single molecule. Even when the cluster is not 
isolated, highly correlated rotors 
still can be expected to manifest themselves 
very differently in reciprocal space
than randomly independent rotors, e.g. in 
a linear chain anticorrelated 
{\em configurations} produced by the dynamics would manifest
themselves  at the $Q$-value corresponding to 
the zone boundary of the reciprocal lattice.
Indeed, the correlated dynamics can give rise 
to static correlations between the orientations
of the molecules, that should be observable.
How is our result then possible?
This question resumes one of the counter-intuitive 
aspects of our result.

The snag is that the clusters we consider can be 
centered on any point of the
crystal lattice, such that two clusters situated 
on neighbouring points can overlap.
If the cog-wheels of two such points have ground 
in succession then the rotations
involved in the second move will partly undo 
the static correlations that existed between the molecules
after the first move. 
In such kind of dynamics the correlations are no longer 
as all-out that we can call them ``ferro''
or ``anti-ferro'' (as in the example of anti-correlated
configurations above). To complicate things further, 
they are not necessarily zero eighter.
They can be of an intermediate strength.
How the story unwinds further, really will depend 
on the specific dynamics
defined within the cog-wheel mechanism.
E.g. it is conceivable that in certain dynamical models
 the memory of the initial
static correlations in a single cluster (or in 
the crystal),  eventually are  totally washed out with time,
due to the game of overlapping clusters.
We then have a different situation than the one
evoked for a problem of an isolated cluster where
the static correlations might persist indefinitely.
Equally conceivable are other models, where such a total loss of memory 
does not occur.

We will distinguish  two 
mutually exclusive possibilities: (1) The loss of 
memory with time is total, such that
starting from any initial configuration that one 
could obtain by independent dynamics
we can stray to any other configuration that 
could be obtained by independent dynamics.
We will call this the ergodic case.
(2) On the contrary, in the non-ergodic case 
not all configurations can be reached. 
The configuration space that applied for the 
independent dynamics then splits up
in several disjoint subsets. Only one of these 
is the set of possible states of the sample
under study. All other states just do not 
exist in the sample.\footnote{Classifying the system according 
to this distinction was
not motivated by the physics. This choice mainly reflects 
the limits of our abilities
to perform certain calculations.
The distinction we make
codes information about what happens to the system at infinite times:
The correlations are eventually lost or otherwise.
Nothing is stated about correlations that might persist
for finite times. Despite this lack of information about
the situation at finite times, in case (1) 
we will be able to infer an  impossibility
to discern indications for the presence of correlations
within the dynamical signal.}

Neutron scattering calculations involve an average over all possible
initial states. In the ergodic case, this averaging 
will run over the same set of initial
states
as in the case of independent dynamics. In the non-ergodic case,
the set is smaller. We are able to prove 
that this kind of averaging over 
all possible configurations
of the molecules in the ergodic case is able to iron out 
the signature of a correlation that was present in the dynamics
of an isolated cluster in reciprocal 
space, i.e. we recover the result
for a single molecule.
In the non-ergodic case, we average over a 
smaller set of possible states.
This kind of weaker averaging does not permit us to assert
that we would recover the result for a single molecule.
That is the result of our paper.

At this point it is perhaps instrumental to summarize
the situation at the hand of an example.
Let us take a linear chain 
whose 2N molecules can take only two
orientations (which we can call pseudospins). 
We suppose that
the dynamics is given by some rule that two 
neighbour sites
can flip simultaneously their pseudospins 
(by two-fold rotations).
If this rule defines permanent dimers in the sense
that only molecules 1+2, 3+4, 5+6, etc. can turn simultaneously,
then we are in the presence of a strong correlation.
The presence of such dimers is immediately obvious
on an instaneous  photograph of the system
and will leave an unmistakeable fingerprint
in the dynamics. That is the case we are all familiar with.
And it is on this case that we all built our intuition.

We render the situation far more subtle by allowing the dimers to be
dissociated by the game of overlapping clusters.\footnote{There is then no longer 
equivalence between the concepts
of a dimer and 
of a (dynamical) cluster as it is used by us in this paper. 
We use the terminology ``dynamical cluster'' to
indicate the presence of a dynamical correlation,
and to define sets of molecules that are supposed 
to be able to turn
together, while 
the terminology ``dimer'' 
indicates the presence of a fixed orientational
relationship between two molecules. Our use
of the word ``cluster'' has also nothing to do
with the idea of a domain wherein all molecules would
have some special orientational relationships, e.g. to
be aligned in a parallel fashion as in a ferromagnet.
Our definition of the word cluster is a purely mathematical one.
It is thus different
from the usual intuitive physical idea.}
Let us first consider a simultaneous 
flip of the molecules on 
sites 1 and 2 (starting from
a completely aligned configuration). 
That introduces a correlation. Now assume 
that an analogous
flip
involving sites 2 and 3 follows. Will  
this remove the ``correlation'', as the ``image'' of the
initial ``dimer'' has been destroyed?
The answer is (here) that some less conspicuous correlation remains.
Due to the construction of the model there will always be
an even number of flipped pseudospins. 
How weak it ever might be, this is a (static) correlation 
that will never
disappear from the model. 
In fact, due to the persistence of these static 
correlations,
the system is not ergodic and does not fulfil the 
assumptions that are underlying our paper.
But in this very specific case it happens  that we can actually 
prove that the model
does {\em not} yield the same neutron answer as a system
with totally uncorrelated molecules. One can easily check
this by using our method of calculation on a small toy model
and exploiting the fact that the configuration space
for the uncorrelated case splits into two disjoint
subsets when we introduce the dynamical correlations.
But if in another type of dynamical model the configuration 
space were not to split,
then we will be able to assert with certainty that
the dynamical signal will not offer clues as to
the presence of the dynamical correlations.

\section{Motivation of the use of our method}

A many reader might find the mathematical leap we will
take to configuration
space too cumbersome and too much of a fuss to his taste. 
Is all this 
really necessary? Part of the answer may reside in the following
warning against a possible confusion.
Consider the following (fallacious) argument.
When we look at an instantaneous photograph of
an ergodic system,  the orientations of the molecules
will in general look completely random.
From this observation it takes only one step
to argue that the ``correlations are random''.
Would that not show that our result is trivial?\cite{Hone}
The problem is that the word ``correlation'' occurs
with several different meanings in this context.
The (spatial) {\em static} correlations between 
the orientations of the molecules
that would be visible
in an instantaneous photograph of the sample may well be 
random.\footnote{In diffraction studies static correlations lead to diffuse
scattering.\cite{Fouret} It is perhaps tempting to invert this implication
in the sense that absence of static correlations implies
absence of diffuse scattering, but this is just wrong in dynamics.}
However, the instant {\em dynamical} (or temporal) correlations
are not random, as they tie up a well-defined
set of molecules into a cog-wheel.

To avoid confusion, it is thus important to realize 
that the word correlation can 
intervene with two entirely different meanings
in the discussion and in the mind of the reader.
We have
{\em static} orientational correlations between 
the molecules of the whole crystal
that we may detect in a snapshot of the system. 
We call such snapshots  configurations.
These static correlations are the ones that matter in 
structural diffraction techniques.
One will run headlong into trouble if one does not
properly sort out for oneself that
moving to  a dynamical context involves a shift of paradigm:
We have to consider {\em temporal} correlations between {\em pairs} of such
snapshots at {\em all} possible times. 
The cog-wheel mechanism defines
{\em dynamical} correlations between molecules 
within a (small) cluster,
in the sense that it tells you which molecules 
will turn simultaneously
and how they will turn. The dynamical 
correlations are the ones that really count,
not the static ones, although the former
are of course responsible
for the occurrence of the latter. In view of 
this remark, the loose phrasing ``in the presence 
of correlations''
appears thus as fraught with ambiguity.\footnote{In this language of static 
and dynamical correlations,
our statement that we do not consider 
environment-dependent dynamics,
can be reformulated by saying that the jump times 
are not considered to depend
on details in the static correlations between 
the orientations of the molecules
inside or outside the dynamical cluster 
within a given configuration, but 
rather uniquely defined
by the cog-wheel mechanism on this dynamical cluster.}

An important aspect of our method is that it develops
a precise mathematical tool that permits to discuss all kinds of
details that can arise in an argument about 
correlations in a very
clear and rigorous way, without running into difficulties 
due to certain inadequacies of plain language.
The examples of the correlations that persist on a linear chain
despite the dissociation of the dimers and of the fallacy
based on the ambiguous use of the phrasing 
``random correlation'' as sketched above 
may serve to show how heedless it is to 
underestimate the difficulty of deciding on 
one's mere intuition
if a correlation is present within a system or otherwise,
and how discussing such matters in informal 
language can seed confusion.
Two persons may easily use the same words and sentences
to refer to mental pictures in their heads that are
fundamentally different, without being aware of these differences. 
In a casual discussion it might e.g.
take quite some time before it transpires that a person does
not take properly into account the vital
distinction between static and dynamical correlations
we outlined above.
The reverse of the medal is that mathematical rigor
all too often leads to austere if not frustrated reading.

\section{Possible domains of application of this work}

\subsection{Scope of experimental probes}

We are presenting our work
in terms of a coherent neutron scattering problem, but
the formalism will also apply to other microscopic techniques,
provided the scattering is really coherent
and the technique allows to obtain a $Q$-dependence.
Coherent scattering does not only occur
with neutrons when the nuclei have an important coherent scattering
cross-section. In contrast with neutron scattering,
where the nuclei can have both a coherent and an incoherent
scattering cross section, X-ray and light scattering processes
jiggle  the electrons and will {\em uniquely} lead to coherent scattering. 
Good candidates  for applications are thus also quasielastic X-ray scattering
and light scattering. In such techniques the $Q$-range 
tends to be small ($Q \approx 0$), but a lot of progress has been made
in the last decade in inelastic X-ray scattering techniques
using synchrotron radiation, rendering available large $Q$-values
combined with good energy resolution ($\approx$ 1 meV).\cite{Sette}
When there is no energy information,
the  energy-integrated quasielastic scattering 
will still yield the ${\mathbf{Q}}$-dependence of the 
diffuse scattering.\footnote{It may be noted that the 
formalism for inelastic X-ray scattering
leads to expressions that link the scattering functions
to the measured intensities in a way that looks identical to 
what we have 
for inelastic neutron scattering. However, important
differences  are very often passed
under silence: The scattering functions that intervene in
X-ray scattering are sums over all the {\em electrons}
in the sample, while in neutron scattering they are sums over
all {\em nuclei}. But both are generally noted
as $S({\bf{Q}}, \omega)$, without any explicit reference to 
this important distinction. Thus some work has 
to be done to render the two different definitions
 of scattering function comparable.
It is not a priori given that the electrons will just
move as the nuclei in the photon field. 
In Raman scattering e.g. the polarizability
of the atom intervenes and gives rise to a coupling factor
that can be hard to evaluate and is different from the simple expression 
${\mathbf{u\cdot Q}}$
that occurs for neutron scattering. (Here ${\mathbf{u}}$ is the atomic
displacement).
The polarization of the electromagnetic radiation is important, etc...
To our knowledge these points have never been worked out exhaustively
for X-ray scattering.}

Extremely high resolutions can be obtained with the 
so-called speckle technique which allows one to study
relaxation on a time scale of the order of 15 minutes!\cite{speckle}
Very good resolutions (neV) can also be obtained by
nuclear resonance methods using synchrotron radiation\cite{Rueffer}
(e.g. exciting the 14.41 keV M\"ossbauer level
of $^{57}$Fe), but in general this is an incoherent process
(as in this process the radiation interacts with the nuclei rather than
with the electrons).
Under certain exeperimental conditions the process is nonetheless coherent,
but then the quasielastic scattering
has to be deteced against a very strong 
background of elastic scattering.

\subsection{Scope of experimental phenomena}

Coherent quasielastic neutron scattering studies are rare
and this is certainly due to the difficulty of treating
the problem theoretically. Incoherent quasielastic neutron scattering
can often be modeled by a system of coupled rate equations,
that one can solve analytically. But describing correlations,
as probed by coherent quasielastic scattering, along such lines 
proves difficult. One solution to this problem is to drop
the ideal of an analytical description all together
and to make Monte Carlo \cite{RossMC} or molecular dynamics 
\cite{moldyn} simulations,
but then fitting some variable parameters of the model 
to the experimental data
will no longer be possible.

There are two main categories of stochastic motion
that lead to (non-magnetic) quasielastic scattering,
viz. translational diffusion and rotational 
relaxation.\footnote{In molecular liquids both occur simultaneously 
and they can be coupled.\cite{Sears}}
Our paper does not deal with translational 
diffusion.\footnote{There is an 
unsuperable intrinsic difference
between rotational and translational diffusion. 
The point is easily made by considering
the translational jump diffusion of two particles
on a linear chain (of $n$ positions). If the two particles could both
occupy a same lattice site simultaneously 
then the configuration space would be just a square lattice
(whose points (1,1),(1,2),...,(i,j),...(n,n), correspond
to configurations where the first particle is located at the site 
with coordinate $i$, and 
the second particle at the site with coordinate $j$).
As real particles cannot be in the same site at the same time,
the diagonal of the square lattice ($i=j$) represents forbidden
configurations. Once these points have been removed
from the square lattice, we are left with two 
totally disconnected mirror-symmetric
triangles (the mirror-symmetry corresponds to 
identical particle exchange). 
Eighter of these triangles can be taken as the
configuration space for the two particles. The generalization
to $N$ particles leads to an $N$-dimensional simplex that
is a $(1/N!)$-th slice of an $N$-dimensional hypercube.
The secular matrix that intervenes in the rate equations 
no longer has $N$-dimensional translational invariance
in this truncated configuration space.
The method of calculation we use in our paper is based 
on $N$-dimensional translational symmetry in configuration space 
and breaks thus down. (In models of correlated rotational jumps
there are no such hyperplane cuts through configuration space 
corresponding to problems of excluded volume,
but it is also a loss of global translational
symmetry, splitting the configuration space
in totally disconnected parts, that renders it impossible to
perform the calculations for the non-ergodic case ).
If it were possible to overcome such 
fundamental difficulties one could
make a mathematically rigorous calculation of the
total scattering function for a hard-sphere
liquid, by taking a continuum limit of an appropriate lattice
gas model.} 
There exists a generalization of the Chudley-Eliott model
that permits to calculate the coherent quasielastic 
neutron scattering  signal
for translational jump diffusion on a lattice.
But this is an average description in that it 
is based on rate equations which rely on the
assumption that the concentrations are rigorously homogeneous
throughout the sample.\cite{Kaisermayr}
 For the sake of completeness
we refer the reader interested in theoretical aspects of 
translational diffusion
to references \cite{RossMC} and \cite{Schroeder}.
The bulk of the experimental efforts have 
focused on the diffusion of 
D in NbD$_{x}$ \cite{NbD},  Ag in silver 
halides \cite{Ag-halides} and in Ag$_{2}$Se 
\cite{Ag2Se}, Rb in RbAg$_{4}$I$_{5}$ \cite{Rb-comp}, 
Na in Na$_{3}$PO$_{4}$ \cite{Na3PO4}
Oxygen in UO$_{2}$ \cite{Fak} and F in fluorites \cite{fluorites}.

What kinds of rotational relaxation is our paper then able to address?
Let us immediately state that our model may not have
a perfect counterpart in the real 
world.\footnote{Eventually the assumptions that underly our model 
are dictated by the limits
of our mathematical skills rather than the physics
we attempted to address. As we will see, our approach of the problem
leads to a description of the dynamics in terms
of huge jump matrices. Analytical
diagonalization of matrices of rank larger than 4 is
in general not possible: As Galois has shown, beyond degree 4
there are no general solutions in terms of radicals 
for polynomial equations.
A notable simplification can occur however,
if we can make use of symmetry. 
It is mainly translational symmetry in configuration space
that achieves the analytical diagonalization 
of the jump matrix in the present paper.}
It is a
requirement of not breaking
translational symmetry in configuration space that imposes the 
underlying assumptions of our paper.
If these are not met,
the calculation of the  correlations
is of a difficulty far beyond the methods we use.

We noted at the beginning of the Introduction 
that the rotating molecules do not need to 
have the symmetry of
the rotational jump. This is important. If the 
symmetry of the molecules
is the same as the symmetry of the jump, 
then the sample will
look the same after the jump as before the jump.
This means that the Fourier transform of the molecules can be
factorized out in the calculations and we do not need to
go through the more difficult approach of our 
paper.\footnote{Unless the dynamics is more complicated
than a simple rotational jump model,
due to supplementary problems such as translation-rotation
coupling\cite{CBr4,CBr4diffuse,t-rcouplingCBr4} etc...}. 

Combining these two limitations of feasibility
and technical interest,  the geometry of the 
nonadecane molecules 
is intrumental for visualizing
the problem that we can treat, and for describing 
the (restricted) kind of correlations that
we have in mind.
The physical correlations  between the nonadecane molecules 
that are really experimentally observed
 in the urea inclusion
compound are  
of a different nature\cite{Toudic2}, 
which we are unable to treat (although this was not yet clear
at the time we started this work).
But this kind of correlations could exist in another system.
The major lesson from our exercise must be that
we give a non-trivial counter-example that indicates how 
coherent quasielastic neutron scattering
may not always yield the information one expects 
on the basis of commonly accepted notions.

In our method the configuration space for the single-molecule 
dynamics is one-dimensional. It is however possible
to generalize the formalism such as to recover
also higher-dimensional single-molecule 
configuration spaces.\footnote{We 
will not give this derivation in order not to burden
the paper. The proof runs as for the one-dimensional case
{\em mutatis mutandis}.} This broadens the scope
of possible applications further to three-dimensional
rotations of molecules whose symmetry is not the same 
as that of the rotation symmetry.
An example of such cases would be the rotational jumps
of molecules of tetrahedral symmetry containing different isotopes
(such that the symmetry is broken)
as e.g. CH$_{3}$D, CD$_{3}$H, CH$_{2}$D$_{2}$, NH$_{3}$D, 
C$^{79}$Br$_{3}$$^{81}$Br, C$^{79}$Br$_{2}$$^{81}$Br$_{2}$ \cite{CBr4}, 
CBr$_{2}$Cl$_{2}$ \cite{CBr2Cl2+CBrCl3}, CBrCl$_{3}$\cite{CBr2Cl2+CBrCl3},
etc...
However in the physical world these molecules do not necessarily give rise
to solids where our type of problems are encountered.
In partially deuterated methane the physics is just different:
Bulk CH$_{3}$D and CD$_{3}$H are either close to rotational diffusion,
or there is quantum behaviour.
In partially deuterated ammonia, the existence of exchange of
$D$ and $H$ between the molecules complicates the picture.

Also octahedral molecules could be considered.\footnote{A good example
would be SF$_{6}$,
if it were not that the natural element
fluorine is mono-isotopic ($^{19}$F), such that the simplification mentioned
above can be applied.\cite{CBr4diffuse}} 
Finally, an important application of our result 
could be C$_{60}$. The (coherent)
quasielastic neutron scattering signal 
of the rotational dynamics
of this molecular crystal in its room 
temperature phase has been 
successfully described in terms 
of a model of completely independent rotors by 
Neumann et al.\cite{Neumann}. Although 
our results have  been
derived for rotational jumps rather 
than for continuous rotations 
as occur in C$_{60}$, they 
suggest by analogy that 
it cannot be claimed with certainty  
on the basis of the
sole neutron data of Neumann 
{\em et al.}  that the molecules would 
be totally independent.

\section{Method}

The general method we use has been already 
described previously \cite{PRB,JdP}. The idea is to formulate
the problem in configuration space. The instantaneous configuration
of the whole system is an abstract particle that diffuses
on a network in this configuration space. The vertices of the network
are the possible configurations of the system.
(How to define them is illustrated by our example of the
translational diffusion of two particles 
in footnote 7). 
When one single particle of
the system makes a jump with a relaxation time $\tau$, we say that
the system ``jumps'' between two configurations with the relaxation time 
$\tau$. Two configurations that are linked by such
a move of one particle are connected by a line
of the network carrying a label $\tau$.
The dynamics of our system is this way mapped isomorphically onto the 
problem  of the diffusion of a
single abstract particle
(the system) on a network in configuration 
space. The only
difference is that the embedding configuration space
is of considerably higher dimension than physical space.

The physical problem can thus be 
described in terms of a set
of coupled rate equations ${d\over{dt}}\,{\mathbf{P}} = {1\over{\tau}}\,
{\mathbf{M}}{\mathbf{P}}$,
where the column vector ${\mathbf{P}}(t)$ contains the probabilities 
$p_{j}(t)$
that the system is in configuration $j$ at time $t$, $\tau$
 is a relaxation time, and
${\mathbf{M}}$ is the so-called jump matrix.
The formal solution is worked out in reference \cite{PRB}
and does not contain further suprising steps or new ideas: 
By diagonalizing 
${\mathbf{M}} =
{\mathbf{S}}\,{\mathbf{\Lambda}}\,{\mathbf{S}}^{-1}$, 
we can write the solution 
of this set of coupled linear differential equations as
${\mathbf{P}}(t) = {\mathbf{S}}\,\exp\,({\mathbf{\Lambda}}t/\tau)\,
{\mathbf{S}}^{-1}\,{\mathbf{P}}(0)$.
By plugging in the various initial conditions into ${\mathbf{P}}(0)$, 
we obtain then the full set
of probabilities $p_{j,k}(t)$ that the system is in configuration $j$ 
at time $t$
if it was in configuration $k$ at time $0$. These are the quantities 
that are needed
in the Van Hove correlation functions. 
A proper choice of initial conditions will also
take into account the thermal occupation factors,
i.e. entail the required thermal averaging.
 The total scattering function is
obtained from the spatio-temporal
Fourier transform of the thermal average of the correlation functions.
In its final expression the total scattering function 
reads: $S({\mathbf{Q}},\omega) = 
{1\over{n^{N}}}\,{\mathbf{F}}\,{\mathbf{S}}\,\,
{\cal{F}}(\,e^{{\mathbf{\Lambda}}t/\tau}\,)\,\,
\,{\mathbf{S}}^{\dag}\,{\mathbf{F}}^{\dag}$,
where ${\mathbf{F}}$ is the  row matrix that 
contains the spatial Fourier transforms $F_{\mathbf{c}}$ of
the configurations ${\mathbf{c}}$ obtained by putting a Dirac measure
of weight $b_{x}$ at the position of each atom of type $x$
($b_{x}$ is its coherent scattering length); ${\cal{F}}$ stands for the
 temporal Fourier transform.
The  normalization factor ${1\over{n^{N}}}$ is here written
for the case of uncorrelated $n$-fold rotational jumps of 
$N$ molecules.\footnote{Generally speaking, treating 
the system in terms of 
configurations has several 
advantages.
First of all, the total scattering function is invariant under the 
operation of identical-particle exchange,
and this fact is automatically accounted for in our 
method, by considering as strictly identical
all the configurations that can be obtained one 
from another by swapping identical particles.
Consequently, identical particles are not tagged 
in our method: They remain undistinguishable.
We  do not
loose  time or efforts in enumerating tagged configurations.
 This reduces the number of configurations
to be considered drastically: For each type 
of particles $j$, there can be a reduction by
a factor $n_{j}!$, where $n_{j}$ stands for the number 
of such identical particles of type $j$ 
(provided that they can indeed physically
swapped by the dynamics). This is e.g. the reason
why in the example of the translational diffusion of two particles
in footnote 7, the configuration space
can be taken as only one of the two triangles defined by the diagonal.
(In the present case of rotational jumps, this simplification does
not occur as the dynamics do not entail exchanges of molecules). 
The second advantage is that all  static correlations that
might occur inside a single configuration are automatically
taken into account.
In the calculation this transpires through the presence of
the Fourier transform ${\cal{F}}_{{\mathbf{c}}}$ of 
the configuration
${\mathbf{c}}$. For this reason we further will 
never even mention these static correlations
in the paper. The worth correlation only occurs 
in a dynamical sense.}

\section{Formulation of the jump matrix}

When there are no correlations
 between the molecules, the configuration space for the 
 $n$-fold rotational
  jumps of $N$ molecules on a
lattice ${\cal{L}} \subset {\mathbb{R}}^{d}\, (d\in \{1,2,3\})$,
 e.g. ${\cal{L}} = (\,[1,\ell]\cap{\mathbb{N}}\,)^{d}$, 
 $N = \ell^{d}$, will be  a
  hypercubic lattice\cite{modulo} 
${\cal{H}} = (\,{\mathbb{Z}}/n\,)^{N}$ with cyclic 
boundary conditions 
in ${\mathbb{R}}^{N}$. 
(Each particle adds one dimension to configuration space:
E.g. if there were just two molecules ($N=2$) that  make
threefold rotational jumps ($n=3$), the configuration
space would be the square lattice of the $n^{N}=3^{2}=9$ points
(1,1), (1,2),...(i,j),... (3,3). The configuration $(i,j)$
corresponds then to the situation that particle 1 has orientation
$i$ and particle 2 orientation $j$. 
The boundary conditions are cyclic 
as each particle
can go through the successive orientations 
$1\rightarrow 2 \rightarrow 3 \rightarrow 1$).
For a three-dimensional sample ($d=3$), the position 
vectors of the molecules will be
 $(\,j_{x},j_{y},j_{z}\,) \in {\cal{L}}$.
The jump matrix ${\mathbf{M}}$ is then defined by:
  
\begin{equation} \label{Eq0}
M_{{\mathbf{c}};{\mathbf{d}}} = -2N \,\delta_{{\mathbf{c}};{\mathbf{d}}} +
 \sum_{{\mathbf{j}}\,\in\,{\cal{L}}}\,
(\,\delta_{{\mathbf{c}};{\mathbf{d}}+{\mathbf{e}}_{{\mathbf{j}}}} + 
\delta_{{\mathbf{c}};{\mathbf{d}}-{\mathbf{e}}_{{\mathbf{j}}}}\,).
\end{equation}

\noindent The reader should not feel intmidated by
the few very concise notations of this type that occur in the paper.
They are mainly given for the sake of completeness.
The main point he should capture is at which points in the
jump matrix there are non-zero entries and how they look like.
Eq. (\ref{Eq0}) is built up as follows.
 There is a unit vector ${\mathbf{e}}_{\,{\mathbf{j}}}$
 in hypercubic space associated with
  each molecule (with position vector ${\mathbf{j}}$) that can turn:
  As we illustrated with the example of two particles above,
  each molecule adds a dimension to  configuration space.
When we start from a configuration ${\mathbf{c}}$ and the molecule ${\mathbf{j}}$
makes a rotational jump, the system goes to the configuration 
${\mathbf{d}} = {\mathbf{c}} +
{\mathbf{e}}_{\,{\mathbf{j}}}$. Hence,
in the absence of correlations, 
 a configuration ${\mathbf{c}} \in {\cal{H}}$ has $2N$ neighbours
 ${\mathbf{d}} = {\mathbf{c}} \pm {\mathbf{e}}_{\,{\mathbf{j}}}$,
  where ${\mathbf{j}} = 
 (\,j_{x},j_{y},j_{z}\,)
 \in {\cal{L}}$: 
At the line corresponding to ${\mathbf{c}}$
 in the jump matrix
there are thus non-zero entries at all columnar positions corresponding
to the configurations
${\mathbf{d}} = {\mathbf{c}} + {\mathbf{e}}_{\,{\mathbf{j}}}$ and 
${\mathbf{d}} = {\mathbf{c}} - {\mathbf{e}}_{\,{\mathbf{j}}}$, 
a fact which is expressed
through the presence of the Kronecker delta's 
$\delta_{{\mathbf{c}};{\mathbf{d}}+{\mathbf{e}}_{{\mathbf{j}}}}$
and $\delta_{{\mathbf{c}};{\mathbf{d}}-{\mathbf{e}}_{{\mathbf{j}}}}$.
The diagonal terms follow from these terms as in any other
jump matrix for a diffusion problem.

 The unit vectors ${\mathbf{e}}_{\,{\mathbf{j}}}$ are the 
generators of the hypercubic 
 lattice ${\cal{H}}$,
 and each configuration ${\mathbf{c}}$ accessible can be written as
 ${\mathbf{c}} = \sum_{\,{\mathbf{j}}}\, c_{\,{\mathbf{j}}}\,
 {\mathbf{e}}_{\,{\mathbf{j}}}$, 
 with $c_{\,{\mathbf{j}}} \in {\mathbb{Z}}/n$.
 The hypercubic lattice ${\cal{H}}$ represents all 
 possible configurations that
 can be obtained by $n$-fold rotations, and therefore 
 the configuration space
 when there are correlations will be a sublattice 
 ${\cal{S}} \subset {\cal{H}}$
  of 
${\cal{H}}$.\footnote{We would like to stress that there is no underlying assumption related
to representing the various configurations as points on a 
hypercubic lattice,
for the mere convenience of enumerating them more easily.
This also true when the rotors are not independent.
More specifically, it does not imply any factorization of probabilities
as we would have in the case of independent rotors.
The probabilities will be given by the topological 
connectivity of the network 
or graph that links these points. These lines could be very different
from the edges of the hypercubes of the hypercubic lattice that occur
in the case of independent dynamics.
The probability for a jump between two configurations 
is taken into account 
by connecting the points  that represent them by a line 
labeled with its relaxation
time. Only if these probabilities themselves were factorized
would we have an underlying independence.}

It is logical that we assume that due to the range of the 
intermolecular interactions,
the set or cluster of jumping molecules extends
to a  neighbour shell of a certain order. Due to the translational
invariance on the physical lattice ${\cal{L}}$,
this cluster should be allowed to occur at every lattice 
site ${\mathbf{j}} \in {\cal{L}}$.
A configuration ${\mathbf{c}}$ will now be connected to other configurations
${\mathbf{c}} \pm {\mathbf{v}}_{{\mathbf{j}}}, {\mathbf{j}} \in  {\cal{L}}$
on the sublattice  ${\cal{S}}$, 
than ${\mathbf{c}} \pm {\mathbf{e}}_{{\mathbf{j}}}, 
{\mathbf{j}} \in  {\cal{L}}$ 
as for the independent dynamics,
and it is the set of $\nu$ relative position vectors
${\mathbf{v}}_{{\mathbf{j}}}$ that has to be defined.
E.g. if on a two-dimensional square lattice ${\cal{L}} =
(\,[1,\ell] \cap {\mathbb{N}}\,)^{2}$
with cyclic boundary conditions, the jump
of a molecule at $(\,j_{x},j_{y}\,)$ over $2\pi/n$ is always accompanied
by opposite jumps over $- 2\pi/n$ of its four first neighbours, 
then $\nu = \ell^{2}$ and

\begin{equation} \label{Eq1}
{\mathbf{v}}_{\,(j_{x},j_{y})} = {\mathbf{e}}_{\,(j_{x},j_{y})}
 - {\mathbf{e}}_{\,(j_{x}+1,j_{y})} - {\mathbf{e}}_{\,(j_{x}-1,j_{y})} 
 - {\mathbf{e}}_{\,(j_{x},j_{y}+1)} - {\mathbf{e}}_{\,(j_{x},j_{y}-1)},
   \forall (\,j_{x},j_{y}\,) \in {\cal{L}},
\end{equation}

\noindent where the minus signs translate the fact that the 
rotations are opposite. We can understand this by decomposing 
the combined move mentally into a succession of single-molecule jumps:
We first turn the molecule $(j_{x},j_{y})$, which changes the
configuration from ${\mathbf{c}}$ to ${\mathbf{c}} + {\mathbf{e}}_{\,(j_{x},j_{y})}$.
Then we turn the molecule $(j_{x}+1,j_{y})$, which changes the configuration
further from ${\mathbf{c}} + {\mathbf{e}}_{\,(j_{x},j_{y})}$ to 
${\mathbf{c}} + {\mathbf{e}}_{\,(j_{x},j_{y})} - {\mathbf{e}}_{\,(j_{x}+1,j_{y})}$,
etc... upto the final configuration ${\mathbf{c}} + {\mathbf{v}}_{\,(j_{x},j_{y})}$.
The order in which we take these individual moves is not important,
as in the combined move of the five molecules the system
does not visit the intermediate configurations and goes
immediately from the initial (${\mathbf{c}}$)
to the final configuration (${\mathbf{c}} + {\mathbf{v}}_{\,(j_{x},j_{y})}$).
On our network there will be a line that connects these two configurations
and which will be labeled by $\tau$. 
{\em Of course the relaxation time $\tau$  we use 
now has  no longer
any relationship whatsoever with the relaxation 
times we used in the problem
of independent dynamics}. It is the relaxation time 
for a cluster, not for a molecule,
and the probabilities we are dealing with here can in principle 
not be expressed as a product of
probabilities as would occur if the rotors were 
independent.\footnote{It must be obvious that the occurrence of a cluster is
probabilistic: The model of Eq. (\ref{Eq1}) is perhaps best seen
as part of a larger model wherein e.g. a molecule has some probability to
turn alone, some probability to turn as given by Eq. (\ref{Eq1}),
and  some probabilities to turn even within
other types of clusters (see e.g. Eq. (\ref{Eq8}) below).
One should thus not use the definition of 
Eq. (\ref{Eq1}) to
``prove''   that the whole lattice
should turn simultaneously: when $(j_{x},j_{y})$ turns,
$(j_{x}+1,j_{y})$ turns; then since $(j_{x}+1,j_{y})$ turns,
$(j_{x}+2,j_{y})$ must also turn, etc...
That would be an absurd way of misinterpreting of 
our aims.}\footnote{We must emphasize that further 
on our approach will only work 
since we assume translational invariance
in configuration space ${\cal{H}}$: the types of possible 
rotational jumps
of a molecule (or a cluster of molecules) and their 
respective probabilities do not depend
on the specific orientation
of that molecule or any other actual detail of the  
prevailing configuration of the system.
This would be quite reasonable for the case of alkane 
molecules evoked, but nevertheless
this remark implies that we have not exhausted all possible meanings of the
 concept correlation.
(We are not dealing in the present paper with the additional
complication that the dynamics could also depend on the actual 
local environment
of the molecules. Defining  systems of the latter 
type in a self-consistent way
is already difficult in its own right. We briefly touch upon it
in the Appendix.).}
The $\nu=\ell^{2}$ vectors ${\mathbf{v}}_{{\mathbf{j}}}$,
 ${\mathbf{j}} = (\,1,1\,), (\,1,2\,), \dots (\,\ell,\ell\,)$ are the
 generators of the sublattice ${\cal{S}}$.
Using the hypercubic norm, we see that $\parallel 
{\mathbf{v}}_{\,(j_{x},j_{y})} 
\parallel = 5$.

As a second example, we could also imagine that each molecule at $(j_{x},j_{y})$ has 
(for symmetry reasons) four equivalent
alternatives to jump simultaneously with a single first neighbour.
Then $\nu = 4\, \ell^{2}$ and  $\forall (j_{x},j_{y}) \in {\cal{L}}$:

\begin{eqnarray} \label{Eq2} 
{\mathbf{v}}_{\,(j_{x},j_{y}),1} = 
{\mathbf{e}}_{\,(j_{x},j_{y})} - 
{\mathbf{e}}_{\,(j_{x}+1,j_{y})},\\\nonumber
{\mathbf{v}}_{\,(j_{x},j_{y}),2} = 
{\mathbf{e}}_{\,(j_{x},j_{y})} - 
{\mathbf{e}}_{\,(j_{x}-1,j_{y})},\\\nonumber
{\mathbf{v}}_{\,(j_{x},j_{y}),3} = 
{\mathbf{e}}_{\,(j_{x},j_{y})} - 
{\mathbf{e}}_{\,(j_{x},j_{y}+1)},\\\nonumber
{\mathbf{v}}_{\,(j_{x},j_{y}),4} = 
{\mathbf{e}}_{\,(j_{x},j_{y})} - 
{\mathbf{e}}_{\,(j_{x},j_{y}-1)}.  
\end{eqnarray}

\noindent Here $\parallel {\mathbf{v}}_{\,(j_{x},j_{y}),r} \parallel = 2$ 
(~$r \in \{1,2,3,4\}$~).
We see thus that in the general situation with correlations the jump 
matrix ${\mathbf{M}}$
will be defined by

\begin{equation} \label{Eq00}
M_{{\mathbf{c}};{\mathbf{d}}} = -2N\rho \,
\delta_{{\mathbf{c}};{\mathbf{d}}} +
 \sum_{{\mathbf{j}}\,\in \,{\cal{L}}}\,
\sum_{r=1}^{\rho}\,
(\,\delta_{{\mathbf{c}};{\mathbf{d}}+{\mathbf{v}}_{{\mathbf{j}},r}} + 
\delta_{{\mathbf{c}};{\mathbf{d}}-{\mathbf{v}}_{{\mathbf{j}},r}}\,),
\end{equation}

\noindent where $\rho = \nu/\ell^{d}$ is the number of 
different types $r$ of clusters per lattice site that can turn (i.e.
$\rho=4$ in the example of Eq. (\ref{Eq2})). The definition of  
Eq. (\ref{Eq00})
is analogous to the one in Eq. (\ref{Eq0}) with each vector 
${\mathbf{e}}_{{\mathbf{j}}}$
replaced by $\rho$ vectors ${\mathbf{v}}_{{\mathbf{j}},r}$.
We can more generally assume that the various possibilities labeled by
$r$ lead to different time constants. Drawing in the time 
constants into ${\mathbf{M}}$ the
jump matrix becomes:

\begin{equation} \label{Eq000}
M_{{\mathbf{c}};{\mathbf{d}}} = 
\sum_{r=1}^{\rho}\,{1\over{\tau_{r}}}\,(\,-2N  \,
\delta_{{\mathbf{c}};{\mathbf{d}}}
  + \sum_{{\mathbf{j}}\,\in \,{\cal{L}}}\,
 (\,\delta_{{\mathbf{c}};{\mathbf{d}}+
 {\mathbf{v}}_{{\mathbf{j}},r}} + 
\delta_{{\mathbf{c}};{\mathbf{d}}-
{\mathbf{v}}_{{\mathbf{j}},r}}\,)\,).
\end{equation}

\noindent This notation can be further generalized by 
noticing that $r$ corresponds
to a difference (or translation) vector ${\mathbf{t}} 
\in {\cal{L}}$,
 $\sum_{r} \rightarrow \sum_{{\mathbf{t}} \in {\cal{L}}}$. 
 We must then admit that for many
values of ${\mathbf{t}}$, we will have 
${1\over{\tau_{{\mathbf{t}}}}} = 0$, or replace
${\mathbf{t}} \in {\cal{L}}$ by ${\mathbf{t}} \in 
{\cal{G}}$ where ${\cal{G}}$ 
is a physical cluster:

\begin{equation} \label{Eq0000}
M_{{\mathbf{c}};{\mathbf{d}}} = 
\sum_{{\mathbf{t}} \in {\cal{G}}}\,{1\over{\tau_{{\mathbf{t}}}}}\,
(\,-2N  \,\delta_{{\mathbf{c}};{\mathbf{d}}}
  + \sum_{{\mathbf{j}}\,\in \,{\cal{L}}}\,
 (\,\delta_{{\mathbf{c}};{\mathbf{d}}+
 {\mathbf{v}}_{{\mathbf{j}},{\mathbf{t}}}} + 
\delta_{{\mathbf{c}};{\mathbf{d}}-
{\mathbf{v}}_{{\mathbf{j}},{\mathbf{t}}}}\,)\,).
\end{equation}

We have just given some examples of possible correlations.
In the present paper we will not further insist on writing down a general
 formalism that would cover
all possible cases. First of all, there are too many possibilities.
Secondly, we think that a general abstract formalism in configuration 
space\cite{remark} (which 
typically has a dimension $\approx10^{24}$)  with notations of the
rather elaborate  type that occur 
in Eq. (\ref{Eq00}-\ref{Eq0000}),
 might easily conceal the rather simple ideas behind our method. 
We will therefore rather proceed by illustrating 
the method for examples of the type given, 
in order to familiarize us with its spirit.

\section{The problem of ergodicity}

Since the hypercubic norms of the vectors 
${\mathbf{v}}_{{\mathbf{j}}}$ 
 are by definition larger than $1$
in the presence of correlations,
 one would be inclined
to think that ${\cal{S}}$ will always be a {\em strict} 
subset of ${\cal{H}}$,
which we expressed by saying that our dynamical problem 
is no longer ``ergodic''. The real situation is more subtle.
Taking the possibility of non-ergodicity  
seriously is a difficult task
that we are unable to treat rigorously
in  general.
In the Appendix we develop a physical argument to
put forward the idea that we can relax the condition
of ergodicity to a weaker criterium of {\em local}
(as opposed to {\em global}) ergodicity.
 For the main stream of the paper we will henceforth
assume ergodicity. But despite the fact that we devote here only a few lines
to this assumption, we must stress that ergodicity is a vital issue for the
validity of the  calculations in the rest of the paper.

\section{Solution of the jump model in the ergodic case}

When the system is ergodic,
  we can immediately use the eigenvectors we already 
  established for ${\cal{H}}$ in reference \cite{PRB},
  and we will be able to derive without effort the 
  eigenvalues from the form of
${\mathbf{v}}_{{\mathbf{j}},r}, \,{\mathbf{j}} \in {\cal{L}}$.
Indeed, when the system is ergodic then the jump
matrix in configuration space has translational symmetry
along each of the directions defined by the
$N$ unit vectors ${\mathbf{e}}_{{\mathbf{j}}}$ that span
${\cal{H}}$. The eigenvectors are therefore
just $N$-dimensional Bloch waves, i.e.  
 $n^{N} \times 1$ column matrices 
${\mathbf{V}}^{({\mathbf{k}})}$
defined by $[\,{\mathbf{V}}^{({\mathbf{k}})}\,]_{{\mathbf{c}}} = 
 \exp\,[\, \imath\, \frac{2\pi}{n}\,
 ({\mathbf{k}} - {\mathbf{k}}_{0})\,{\mathbf{\cdot}}\,
 ({\mathbf{c}} - {\mathbf{k}}_{0})\,], \forall {\mathbf{c}} 
 \in {\cal{H}}$.
  Here ${\mathbf{k}}_{0}$ 
 stands for $(1,1,1,\dots1,1) \in {\mathbb{R}}^{N}$
and is introduced to take into account the fact that the expression
for the eigenvectors features the quantities $k_{\xi}$ and $c_{\xi}$
 always under the form of
  linear combinations $(k_{\xi}-1)$, and $(c_{\xi}-1)$.
The $N$-dimensional Bloch wave is obtained
as a Kronecker product of $N$ one-dimensional
Bloch waves, just as a three-dimensional phonon
can also be written as a Kronecker product of three one-dimensional
Bloch-waves. (Each one-dimensional Bloch wave is a $n \times 1$ column matrix).
 Combining these eigenvectors
 with the definition of the $n^{N} \times n^{N}$ jump 
 matrix ${\mathbf{M}}$
  in Eq. (\ref{Eq00}) yields the corresponding eigenvalues:
 
\begin{equation} \label{Eq5a}
\lambda_{\mathbf{k}} = - 2\rho N + 
2\,\sum_{{\mathbf{j}}\,\in\,{\cal{L}}}\,\sum_{r=1}^{\rho}\,
\cos\,[\,{\mathbf{v}}_{{\mathbf{j}},r}\,
{\mathbf{\cdot}}\,(\,{\mathbf{k}} - {\mathbf{k}}_{0}\,)\,].
\end{equation}

\noindent  This is exactly analogous to the way
the eigenvalues of a phonon problem are obtained by
operating the dynamical matrix on the Bloch wave eigenvectors.
We just calculate
${\mathbf{M V}}^{({\mathbf{k}})}$
using the definitions of ${\mathbf{M}}$
and ${\mathbf{V}}^{({\mathbf{k}})}$
and check that the result can be rewritten
as $\lambda^{({\mathbf{k}})}\,
{\mathbf{V}}^{({\mathbf{k}})}$
for some value $\lambda^{({\mathbf{k}})}$.
For the phonon case in text books,  this 
calculation is usually only written for a general
line of the column matrix that represents the eigen vector,
with the possible effect that the argument might not
be recognized as perfectly analogous to the one 
we are dealing with here.
E.g. in the case of Eq. (\ref{Eq1}) we have 
$N=\ell^{2}$, $\rho=1$,
 $\forall {\mathbf{j}} 
\in {\cal{L}} : {\mathbf{k}}_{0}\,
{\mathbf{\cdot\,v}}_{{\mathbf{j}}} = -3$,
 such that Eq. (\ref{Eq5a}) is more explicitly  seen to yield

\begin{eqnarray} \label{Eq5}
\begin{tabular}{lccl}
$\lambda_{\mathbf{k}}$ & $=$ & $- 2\ell^{2}$ &  
$+ 2 \cos\, [\,\frac{2\pi}{n}(\,k_{\,(1,1)} - k_{\,(1,2)} - 
k_{\,(1,\ell)} 
- k_{\,(2,1)} - k_{\,(\ell,1)} + 3\,)\,]$\\\\ 
 & & & $+ 2 \cos \,[\,\frac{2\pi}{n}(\,- k_{\,(1,1)} + 
 k_{\,(1,2)} - k_{\,(1,3)}
 - k_{\,(2,2)} - k_{\,(\ell,2)} + 3\,)\,]$\\
  & & & $\quad\quad\quad\quad\quad\quad\quad\vdots$ \\
 & & & $  + 2 \cos\, [\,\frac{2\pi}{n}(\,k_{\,(j_{x},j_{y})}
 - k_{\,(j_{x}+1,j_{y})} - k_{\,(j_{x}-1,j_{y})} 
 - k_{\,(j_{x},j_{y}+1)} - k_{\,(j_{x},j_{y}-1)} + 3\,)\,] $\\
  & & & $\quad\quad\quad\quad\quad\quad\quad\vdots$\\
 & & & $ + 2 \cos \,[\,\frac{2\pi}{n}\,- k_{\,(1,\ell)} - 
 k_{\,(\ell-1,\ell)} 
 - k_{\,(\ell,1)} - k_{\,(\ell,\ell-1)}
+ k_{\,(\ell,\ell)} + 3\,)\,]$,\\
\end{tabular}
\end{eqnarray}

\noindent where ${\mathbf{k}} = (\,k_{\,(1,1)}, k_{\,(1,2)}, 
\dots k_{\,(j_{x},j_{y})},
\dots k_{\,(\ell,\ell)}\,)$, and
$k_{\,(j_{x},j_{y})} \in {\mathbb{Z}}/n$. There are thus $n^{\ell^{2}}$ 
${\mathbf{k}}$-vectors in the case of Eq. (\ref{Eq1}). 
In fact, a general configuration ${\mathbf{c}}$
has $\ell^{2}$ vector components: ${\mathbf{c}} = (c_{\,(1,1)}, c_{\,(1,2)},
\cdots c_{\,(j_{x},j_{y})},\cdots c_{\,(\ell,\ell)})$. The coordinates
$c_{\,(j_{x},j_{y})}$ can take $n$ values representing the $n$
possible orientations of the molecule at $(j_{x},j_{y}) \in {\cal{L}}$.
There are thus in total $n^{\ell^{2}}$ configurations, and
for each of them there is a probability.
The jump matrix ${\mathbf{M}}$ works on the space of these 
$n^{\ell^{2}}$ configurations. An eigenvector 
${\mathbf{V}}^{({\mathbf{k}})}$ is 
a vector with $n^{\ell^{2}}$ entries
$[\,{\mathbf{V}}^{({\mathbf{k}})}\,]_{\,{\mathbf{c}}}$,
where ${\mathbf{c}}$ runs through all possible configurations.
We can thus depicture ${\mathbf{V}}^{({\mathbf{k}})}$ as a 
function (the Bloch wave):
${\mathbf{V}}^{({\mathbf{k}})}: {\mathbf{c}} \rightarrow 
[\,{\mathbf{V}}^{({\mathbf{k}})}\,]_{\,{\mathbf{c}}}$.
Let us operate ${\mathbf{M}}$ on an eigenvector.
The first non-zero off-diagonal entry in ${\mathbf{M}}$
corresponds to the possibility that ${\mathbf{c}}$ undergoes
a transformation to ${\mathbf{c}} + {\mathbf{v}}_{\,(1,1)}$.
That will transform the function from 
$\exp\,[\, \imath\, \frac{2\pi}{n}\,
 ({\mathbf{k}} - {\mathbf{k}}_{0})\,{\mathbf{\cdot}}\,
 ({\mathbf{c}} - {\mathbf{k}}_{0})\,], \forall {\mathbf{c}} 
 \in {\cal{H}}$ to $\exp\,[\, \imath\, \frac{2\pi}{n}\,
 ({\mathbf{k}} - {\mathbf{k}}_{0})\,{\mathbf{\cdot}}\,
 ({\mathbf{c}} + {\mathbf{v}}_{\,(1,1)} - {\mathbf{k}}_{0})\,], 
 \forall {\mathbf{c}} 
 \in {\cal{H}}$, i.e. multiply it by
 $\exp[\,\imath\frac{2\pi}{n}(\,k_{\,(1,1)} - k_{\,(1,2)} - 
k_{\,(1,\ell)} 
- k_{\,(2,1)} - k_{\,(\ell,1)} + 3\,)\,]$.
Continuing this way with the possible
transformations towards ${\mathbf{c}} + {\mathbf{v}}_{\,(1,2)}$,...,
${\mathbf{c}} + {\mathbf{v}}_{\,(\ell,\ell)}$, including also
the possible transformations  towards
${\mathbf{c}} - {\mathbf{v}}_{\,(1,1)}$,
${\mathbf{c}} - {\mathbf{v}}_{\,(1,2)}$,...,
${\mathbf{c}} - {\mathbf{v}}_{\,(\ell,\ell)}$, and finally adding
the diagonal term
we obtain the result announced in Eq. (\ref{Eq5}).

The whole part of the calculation involving the determination of the
 structure factors remains
the same as for the case without correlations. In fact this calculation 
(based on the evaluation of ${\mathbf{G}} = {\mathbf{F}}\,{\mathbf{S}}$) is entirely
defineded by the values of the eigenvectors (${\mathbf{S}}$) 
and the Fourier transforms of
 the configurations (${\mathbf{F}}$)
and these are not changed in the ergodic case. This means that Eqs.
 (35) and (37)
of reference \cite{PRB} remain valid. Using the same notations
as in that paper,  
the Lorentzian 
$L(\hbar\lambda_{\mathbf{k}}/\tau,\omega)$
will be associated with $|\,G_{\mathbf{k}}\,|^{\,2}$, where

\begin{eqnarray}  \label{Eq6}
%\begin{tabular}{lr} 
G_{k_{\,(1,1)};\,k_{\,(1,2)};\,\dots k_{\,(j_{x},j_{y})};\,
\dots \,k_{\,(\ell,\ell)}} = 
 {1\over{n}}\,\,\sum_{j_{x}=1}^{\ell}\,\sum_{j_{y}=1}^{\ell}\,
 \,\sum_{c_{(j_{x},j_{y})}=1}^{n}\,{\cal{F}}_{c_{(j_{x},j_{y})}}\,
e^{\imath \frac{2\pi}{n}\, (c_{(j_{x},j_{y})}-1)\,(k_{(j_{x},j_{y})}-1)}\,
e^{\imath {\mathbf{Q\cdot r}}_{\,(j_{x},j_{y})}}\,\,\\\nonumber\\\nonumber
\quad\quad\quad\quad\quad\quad\quad\quad\quad\quad\quad\quad\quad\quad\quad
\times \,\,\delta_{1\,k_{\,(1,1)}}\,\,
\delta_{1\,k_{\,(1,2)};\,1}\,\,\dots \,\,
\delta_{1\,k_{\,(j_{x},j_{y}-1)}}\,\,
\delta_{1\,k_{\,(j_{x},j_{y}+1)}}\,\,
\dots\,\,\delta_{1\,k_{\,(\ell,\ell-1)}}\,\,
\delta_{1\,k_{\,(\ell,\ell)}}.
%\end{tabular}
\end{eqnarray}

\noindent Here ${\cal{F}}_{c_{(j_{x},j_{y})}}$ is the Fourier transform
of the  molecule $(j_{x},j_{y})$ (with orientation $c_{(j_{x},j_{y})}$\,)
if it were placed at the origin. (The real position vector 
of this molecule is
 ${\mathbf{r}}_{\,(j_{x},j_{y})}$).
Consequently, the structure factor of the elastic term
remains the same. The structure factors of the quasielastic 
terms remain also the same.
Just the widths of the Lorentzians have to be re-examined.
Due to the presence of the Kronecker symbols in Eq. (\ref{Eq6}) 
only Lorentzians 
for ${\mathbf{k}}$-values which have one
single component $k_{\,(j_{x},j_{y})} \neq 1$
are contributing. In our example on the square lattice, 
these are the Lorentzians with

\begin{equation} \label{Eq7}
\lambda_{\mathbf{k}} = - 2\ell^{2} + 2\,(\ell^{2} - 5) + 
10 \, \cos\, [\,\frac{2\pi}{n}(\,k_{\,(j_{x},j_{y})} -1\,)\,],
\end{equation}

\noindent since in $\lambda_{\mathbf{k}}$ there are 
$\ell^{2} - 5$ terms with
 a vanishing argument
in the cosines and $5$ terms where the argument of the cosines
collapse to $\frac{2\pi}{n}(\,k_{\,(j_{x},j_{y})} -1\,)$.
We recover thus miraculously the same Lorentzians with a 
width parameter
$-2 + 2 \cos\,[\,\frac{2\pi}{n}(\,k_{\,(j_{x},j_{y})} -1\,)\,]$
 as in the case without correlations, except
for the detail that the relaxation time $\tau$ that occurs in 
the rate equations
is being replaced in the final result by a five  times faster
relaxation time $\tau/5$, which does not happen in the case 
without correlations.\footnote{But since we are 
dealing with clusters
of $5$ molecules, we might originally have had the intuition to put a 
value $5$ times slower
than in the uncorrelated case into
the rate equations. If we do not allow for this, then we would  
end up with a speed-up
 of the quasielastic signal
in the presence of reorienting clusters, which is quite unphysical,
 as it should in principle be more difficult to move a set of molecules
 rather than a single molecule.
 This use of a factor $5$ based on a handwaving argument 
 can only be considered
 as a first approximation. We have no means to elucidate 
 this relationship any further, 
 since the jump time a single molecule would have had if 
 its jumps had been independent
 is just not available to the experimentalist.
 There could be a change in the jump time parameter 
 between the cases
 of correlated and uncorrelated jumps, but this cannot 
 be learned
 from an inspection of coherent scattering data alone, 
 since there is no change in the structure
 factors.} {\em There is only a possible renormalization 
 of the relaxation time
 to be used in the description of the jump dynamics.}
 Conclusion: It is absolutely impossible to appreciate 
 from a signal due
to coherent neutron scattering alone if there are 
correlations or otherwise!
One can only hope that a comparison with incoherent 
data might still yield some clues.
This result will be general if the cardinal number 
$\# {\cal{H}}$
of the set of possible configurations
 ${\cal{H}}$
is not altered by the presence of correlations (case (1) in 
the terminology of
the Introduction).
In fact, the factor $5$ in our example comes from the 
circumstance that
our clusters contain five molecules, such that 
${\mathbf{v}}_{\,(j_{x},j_{y})}$ will
contain five terms. Due to this $\lambda_{\mathbf{k}}$ 
will contain five terms
that contain $(\,k_{\,(j_{x},j_{y})}-1\,)$, each time 
combined with other
terms of the general type $\pm (\,k_{\,(j_{x}',j_{y}')}-1\,)$, 
that vanish since
 $k_{\,(j_{x}',j_{y}')} = 1$. All other contributions to 
 $\lambda_{\mathbf{k}}$
are devoid of terms in $k_{\,(j_{x},j_{y})}$.

We have not addressed the possibility illustrated by the 
example of Eq. (\ref{Eq2}),
where we have more than one type of cluster.
In principle this case  is not different in nature from the 
one embodied
by Eq. (\ref{Eq1}), since there we also have already 
$\nu > N$, such that Eq. (\ref{Eq1})
represents two equations rather than one.
 Provided we have ergodicity we can derive
  for  this model from Eq. (\ref{Eq00}) with 
  $N=\ell^{2}, \rho=4$, $(\,\forall {\mathbf{j}} 
\in {\cal{L}}\,) \,(\,\forall r \in 
\{1,2,3,4 \}\,)\,(\,{\mathbf{k}}_{0}\,
{\mathbf{\cdot\,v}}_{{\mathbf{j}},r} = 0\,)$:

\begin{eqnarray} \label{Eq5b}
\begin{tabular}{lccl}
$\lambda_{\mathbf{k}}$ & $=$ & $- 8\ell^{2}$ &   
$+ 2 \cos\, [\,\frac{2\pi}{n}(\,k_{\,(1,1)} - k_{\,(1,2)}\,)\,]
 + 2 \cos\, [\,\frac{2\pi}{n}(\,k_{\,(1,1)} - k_{\,(1,\ell)}\,)\,]$\\\\
  & & &  
$+ 2 \cos\, [\,\frac{2\pi}{n}(\,k_{\,(1,1)} - k_{\,(2,1)}\,)\,]
 + 2 \cos\, [\,\frac{2\pi}{n}(\,k_{\,(1,1)} - k_{\,(\ell,1)}\,)\,]$\\\\ 
  & & & 
$+ 2 \cos\, [\,\frac{2\pi}{n}(\,k_{\,(1,2)} - k_{\,(1,1)}\,)\,]
 + 2 \cos\, [\,\frac{2\pi}{n}(\,k_{\,(1,2)} - k_{\,(1,3)}\,)\,]$\\\\ 
  & & & 
$+ 2 \cos\, [\,\frac{2\pi}{n}(\,k_{\,(1,2)} - k_{\,(2,2)}\,)\,]
 + 2 \cos\, [\,\frac{2\pi}{n}(\,k_{\,(1,2)} - k_{\,(\ell,2)}\,)\,]$\\
 & & & $\quad\quad\quad\quad\quad\quad\quad\vdots$ \\
  & & & 
$+ 2 \cos\, [\,\frac{2\pi}{n}(\,k_{\,(j_{x},j_{y})} - 
k_{\,(j_{x}+1,j_{y})}\,)\,]
 + 2 \cos\, [\,\frac{2\pi}{n}(\,k_{\,(j_{x},j_{y})} - 
 k_{\,(j_{x}-1,j_{y})}\,)\,]$\\\\
  & & & 
$+ 2 \cos\, [\,\frac{2\pi}{n}(\,k_{\,(j_{x},j_{y})} - 
k_{\,(j_{x},j_{y}+1)}\,)\,]
 + 2 \cos\, [\,\frac{2\pi}{n}(\,k_{\,(j_{x},j_{y})} - 
 k_{\,(j_{x},j_{y}-1)}\,)\,]$\\
  & & & 
$\quad\quad\quad\quad\quad\quad\quad\vdots$\\
 & & & 
$+ 2 \cos\, [\,\frac{2\pi}{n}(\,k_{\,(\ell,\ell)} - 
k_{\,(1,\ell})\,)\,]
 + 2 \cos\, [\,\frac{2\pi}{n}(\,k_{\,(\ell,\ell)} - 
 k_{\,(\ell-1,\ell)}\,)\,]$\\\\
  & & &  
$+ 2 \cos\, [\,\frac{2\pi}{n}(\,k_{\,(\ell,\ell)} - 
k_{\,(\ell,1)}\,)\,]
 + 2 \cos\, [\,\frac{2\pi}{n}(\,k_{\,(\ell,\ell)} - 
 k_{\,(\ell,\ell-1)}\,)\,],$\\
\end{tabular} 
\end{eqnarray}

\noindent Expressing the selection rule from Eq. (\ref{Eq6}) that only 
Lorentzians are contributing
 which correspond to a ${\mathbf{k}}$-value that contains a
single component $k_{\,(j_{x},j_{y})} \neq 1$, we just 
retain the Lorentzians with

\begin{equation} \label{Eq7b}
\lambda_{\mathbf{k}} = - 8\ell^{2} + 2\,(4\,\ell^{2} - 8) + 
16 \, \cos\, [\,\frac{2\pi}{n}(\,k_{\,(j_{x},j_{y})} -1\,)\,],
\end{equation}

\noindent such that the widths are now again
 $-2 + 2  \cos\,[\,\frac{2\pi}{n}(\,k_{\,(j_{x},j_{y})} -1\,)\,]$,
this time with a prefactor $8$, since there are in total $16$ 
different moves that involve
a jump of molecule $(\,j_{x},j_{y}\,)$.

We may note that we have already covered quite 
realistically looking possibilities.
The task of taking into account correlations of 
this type on a whole lattice
 definitely  seemed daunting before we embarked 
 on our method.
 The most realistic case would probably involve 
 a  distribution of cluster sizes, 
whereby each cluster has its own characteristic 
relaxation time.
E.g. on the square lattice we could instead of
Eq. (\ref{Eq1}), have relaxation times and clusters 
$\forall (\,j_{x},j_{y}\,) \in {\cal{L}}$:

\begin{eqnarray} \label{Eq8}
\begin{tabular}{llcl}
$\tau_{1}:$ & ${\mathbf{v}}_{\,(j_{x},j_{y}),1}$ & 
$=$ & ${\mathbf{e}}_{\,(j_{x},j_{y})}
 - {\mathbf{e}}_{\,(j_{x}+1,j_{y})} - 
 {\mathbf{e}}_{\,(j_{x}-1,j_{y})} 
 - {\mathbf{e}}_{\,(j_{x},j_{y}+1)} - 
 {\mathbf{e}}_{\,(j_{x},j_{y}-1)}$,
 \\\nonumber\\\nonumber
$\tau_{2}:$ & ${\mathbf{v}}_{\,(j_{x},j_{y}),2}$ & 
$=$ & ${\mathbf{e}}_{\,(j_{x},j_{y})}
 - {\mathbf{e}}_{\,(j_{x}+1,j_{y})} - 
 {\mathbf{e}}_{\,(j_{x}-1,j_{y})} 
 - {\mathbf{e}}_{\,(j_{x},j_{y}+1)} - 
 {\mathbf{e}}_{\,(j_{x},j_{y}-1)}
 + {\mathbf{e}}_{\,(j_{x}-1,j_{y}+1)} 
 + {\mathbf{e}}_{\,(j_{x}+1,j_{y}-1)} 
 $\\\nonumber\\\nonumber
~&~&~& $ \quad\quad\quad +~{\mathbf{e}}_{\,(j_{x}+2,j_{y})} +
 {\mathbf{e}}_{\,(j_{x}-1,j_{y}-1)}
 + {\mathbf{e}}_{\,(j_{x}-2,j_{y})} 
 + {\mathbf{e}}_{\,(j_{x},j_{y}+2)} + 
 {\mathbf{e}}_{\,(j_{x},j_{y}-2)} 
 + {\mathbf{e}}_{\,(j_{x}+1,j_{y}+1)}$,
\end{tabular}\\
\end{eqnarray}

\noindent where in the additional cluster we have 
included now all members 
of the second-neighbour shell,
 assuming that their rotional jumps
are  in phase
with those of the central molecule, i.e. in opposite 
phase with all the members
of the first-neighbour shell.
This leads to Lorentzians with a width parameter
${1\over{\tau_{1}}}\,\lambda_{\mathbf{k}}^{(1)} +
{1\over{\tau_{2}}}\,\lambda_{\mathbf{k}}^{(2)}$, where 
$\lambda_{\mathbf{k}}^{(1)}$ is given by
Eq. (\ref{Eq5}), while $\lambda_{\mathbf{k}}^{(2)} = 
-2\ell^{2} + 2\,\sum_{(j_{x},j_{y})\,\in\,{\cal{L}}}\,
\cos \,[\,\frac{2\pi}{n}\,(\,k_{\,(j_{x},j_{y})}
 - k_{\,(j_{x}+1,j_{y})}
 - k_{\,(j_{x}-1,j_{y})} 
 - k_{\,(j_{x},j_{y}+1)} 
 - k_{\,(j_{x},j_{y}-1)}
 + k_{\,(j_{x}-1,j_{y}+1)} 
 + k_{\,(j_{x}+1,j_{y}-1)}
 + k_{\,(j_{x}+2,j_{y})}
 + k_{\,(j_{x}-1,j_{y}-1)}
 + k_{\,(j_{x}-2,j_{y})} 
 + k_{\,(j_{x},j_{y}+2)}
 + k_{\,(j_{x},j_{y}-2)} 
 + k_{\,(j_{x}+1,j_{y}+1)} 
 -5\,)\,]$. 
 We can appreciate that in general the width parameter for 
 the Lorentzian $L_{\mathbf{k}}$
 will be
 
 \begin{equation} \label{Eq8b}
 \sum_{r=1}^{\rho}\,{1\over{\tau_{r}}}\,(\,-2N + 2\,
 \sum_{ {\mathbf{j}}\,\in\,{\cal{L}} }\,
 \cos \,[\,\frac{2\pi}{n}\,{\cal{P}}_{{\mathbf{j}},r}\,]\,),
 \end{equation}

\noindent since each cluster, i.e. each
 term ${\mathbf{v}}_{\,{\mathbf{j}},r}$ introduces a term
  $-2N + 
 2\,\sum_{ {\mathbf{j}}\,\in\,{\cal{L}} }\,
 \cos\,(\,2\pi\,{\cal{P}}_{{\mathbf{j}},r}/n\,)$,
 where the polynomial ${\cal{P}}_{{\mathbf{j}},r}$ in the 
 $k$'s is obtained by replacing
 each symbol ${\mathbf{e}}$ by a symbol $k$ in the definition of
  ${\mathbf{v}}_{\,{\mathbf{j}},r}$
 and adding $- {\mathbf{k}}_{0}\,
 {\mathbf{\cdot\,v}}_{{\mathbf{j}},r}$.
 After combining with $\delta_{1\,k_{\,(1,1)}}\,\,
 \delta_{1\,k_{\,(1,2)}}\,\,\dots \,\,
\delta_{1\,k_{\,(j_{x},j_{y}-1)}}\,\,
\delta_{1\,k_{\,(j_{x},j_{y}+1)}}\,\,
\dots\,\,\delta_{1\,k_{\,(\ell,\ell-1)}}\,\,
\delta_{1\,k_{\,(\ell,\ell)}}$
in the example of Eq. (\ref{Eq8}) we recover 
 $-4 (\,{5\over{\tau_{1}}}\,+ {13\over{\tau_{2}}}\,)\,
\sin^{2}\,[\,\frac{\pi}{n}(\,k -1\,)\,]$ 
provided ${\cal{S}}={\cal{H}}$.
(Here we have used the fact that the set of values 
$k_{\,(j_{x},j_{y})} \in {\mathbb{Z}}/n$ 
 is the same for all molecules $(j_{x},j_{y})$, 
 such that identical values coming
 from different values of $(j_{x},j_{y})$ can 
 be regrouped in a term where the $k$-value
 is simply noted as $k$).
The more different types of clusters one includes,
the more likely the configuration space ${\cal{S}}$ 
will cover the whole of ${\cal{H}}$.
In fact, if also single-molecule jumps are allowed we
forcedly end up with ${\cal{S}}={\cal{H}}$.
Also for the final expression of the Lorentzian widths after
 the application of the selection rules
it is not difficult to derive the general result:
 $-4\,\,[\,\sum_{r=1}^{\rho}\,(\,m_{r}/\tau_{r}\,)\,]\,
 \sin^{2}\,[\,\frac{\pi}{n}(\,k -1\,)\,]$,
  where $m_{r}$ is the number
of molecules involved in cluster $r$.
The use of the quantity $m_{r}$ exploits the fact that in general
 the correlations between the molecules
will be symmetrical.
The most striking point is that all this follows 
almost effortless, even for
very complex situations, as the translational 
invariance on ${\cal{H}}$ establishes
 a kind of dictionary 
 ${\mathbf{v}}_{\,{\mathbf{j}},r} \rightarrow 
 {\cal{P}}_{{\mathbf{j}},r}$
which allows immediately to write down the eigenvalues, 
and that these eigenvalues
are further tremendously simplified by the selection rule 
for the ${\mathbf{k}}$-values.

\section{Conclusion}

We conclude:
Apart from a possible renormalization of
the jump time, the signal in presence of correlations 
%
% begin added
%
of the type we considered
%
% end added
%
is the same  
as in total absence of correlations.
In other words: Coherent quasielastic neutron scattering 
is unable to reveal possible
correlations 
%
% begin added
%
of the type we considered 
%
% end added
%
between the rotational jumps of the molecules.
This is a clear, but non-trivial result.
%
% removed
%
% , and it does not look plausible to us
% that it could be obtained by some simple, intuitive argument.
%
%
%
We may call it paradoxical that coherent scattering cannot reveal 
%
% begin added
%
the type of correlations we considered,
%
% end added
%
while incoherent scattering can.
%
% begin added
%
The finding that the widths are renormalized calls for caution
in the interpretation of such widths in the results of 
an experiment. One would be inclined
to interpret them as simple quantities corresponding to the relaxation
of a single molecule, while their real meaning could be quite different.
That is another important result of our work.
%
% end added
%

{\em Acknowledgements.} The author would like to thank 
M. Duneau for fruitful discussions.

\section*{Appendix: The problem of ergodicity}

\subsection{Why the problem is not simple}

As the hypercubic norms of the vectors ${\mathbf{v}}_{{\mathbf{j}}}$ 
are by definition larger than $1$
in the presence of correlations, one might believe that we 
never have ergodicity. They span a larger fundamental cell in hyperspace
than the unit cell that occurs with independent dynamics,
and therefore less of such cells should be contained in ${\cal{H}}$.
But since we have cyclic boundary conditions in each dimension
of the hypercubic lattice ${\cal{H}}$ this is not necessarily true. 
A good example to illustrate this point is the transformation 
${\mathbf{v}}_{1}={\mathbf{e}}_{x} + {\mathbf{e}}_{y}$,
${\mathbf{v}}_{2}={\mathbf{e}}_{x} - {\mathbf{e}}_{y}$ on a square
lattice, which splits ${\mathbb{Z}}^{2}$
into two sublattices, one with integer and one with half-integer 
coordinates
in the not normalized basis ${\mathbf{v}}_{1},{\mathbf{v}}_{2}$.
But if we apply this transformation to the restriction 
$({\mathbb{Z}}/n)^{2}$,
allowing for  cyclic boundary
conditions, then for $n$ odd the vectors ${\mathbf{e}}_{1}$ 
and ${\mathbf{e}}_{2}$
will generate the whole lattice. We see thus that the problem of 
ergodicity we ran into
is not as trivial to cope with as one might have thought 
at the start.
It is however very important to know if we can claim that the 
problem is ergodic,
since it determines if the calculation presented in this 
paper applies or otherwise.

Let us consider in the non-ergodic case  a configuration 
${\mathbf{c}}_{1}$ that occurs in ${\cal{H}}$,
but not in the  sublattice
${\cal{S}}$. On this configuration we can apply also the set of
generators ${\mathbf{v}}_{p}, p=1, 2, \dots \nu$, to obtain 
the coset ${\mathbf{c}}_{1} + {\cal{S}}$. The term coset is 
here appropriate
since the lattices ${\cal{H}}$ and ${\cal{S}}$ can be 
considered as visualizations
of a translation group.
There might be a further configuration ${\mathbf{c}}_{2}$ 
that does not occur within
${\cal{S}} \cup ({\mathbf{c}}_{1} + {\cal{S}})$.
This way we construct $s = \# {\cal{H}}/\# {\cal{S}}$ 
disjoint cosets, based
on initial configurations ${\mathbf{c}}_{1}, 
{\mathbf{c}}_{2}, \dots {\mathbf{c}}_{s}$.
Our lattice ${\cal{H}}$ decomposes then as a  
Bravais lattice ${\cal{S}}$
convoluted with a unit cell containing $s$ ``atoms'' 
${\mathbf{c}}_{1}, {\mathbf{c}}_{2},  \dots {\mathbf{c}}_{s}$.

The correlated dynamics will then map onto the diffusion 
of an abstract
particle (the configuration of the system) on the connected 
graph whose vertices
are the points of ${\cal{S}}$. From any point of this graph 
it will be possible
to wander to any other point of this graph.
A point from another coset cannot be reached. Another coset represents
thus a similar sample with similar dynamics, but the states of the sample
represented by this coset cannot be reached starting from the states of the
sample represented by ${\cal{S}}$.  The $s$ cosets represent thus $s$
different samples with a similar structure and similar dynamics,
that would be a kind of ``eniantomorphs''. The different 
samples are characterized
by the fact that they cannot have the same initial state.
It is  {\em a priori} not true that working 
with one of these $s$ enantiomorphs 
should still yield the same ${\bf Q}$-dependence
as for the independent dynamics.
This is shown by the counter example
with the Ising pseudospins given in the Introduction.

The main problem is that we would have to do the thermal averaging over
${\cal{S}}$ rather than over ${\cal{H}}$. We would therefore 
need to determine
how the lattice ${\cal{S}}$ looks like and to find a set of 
basic vectors that
generates it. In general, this will prove just too 
difficult a task,
due to the huge dimensions of the configuration space.
Furthermore, the averaging over ${\cal{S}}$ will in 
general be too weak.
Many sums in the averaging over ${\cal{H}}$ 
reduce to Kronecker deltas (see reference \cite{PRB}). 
In the average over ${\cal{S}}$, 
these sums will
contain less terms, and therefore no longer necessarily yield  
results that are either $0$ or $1$. 

\subsection{Obvious mathematical criterium of non-ergodicity}

One can think of one simple way to find out that the 
sample is not ergodic, viz.
when the dynamics can be described by a single equation,
as e.g. in the example of Eq. (\ref{Eq1})
and the dimension of ${\cal{S}}$ is lower than the 
dimension of ${\cal{H}}$.
There is a linear transformation with matrix ${\mathbf{T}}$:

\begin{equation} \label{Eq3}
[\,{\mathbf{v}}_{\,(1,1)}, {\mathbf{v}}_{\,(1,2)}, 
\dots{\mathbf{v}}_{\,(2,1)},
\dots {\mathbf{v}}_{\,(\ell,\ell)}\,]^{\top} = {\mathbf{T}}\,
[\,{\mathbf{e}}_{\,(1,1)}, {\mathbf{e}}_{\,(1,2)}, 
\dots {\mathbf{e}}_{\,(2,1)},
\dots {\mathbf{e}}_{\,(\ell,\ell)}\,]^{\top}.
\end{equation}

\noindent If $\det{\mathbf{T}} = 0$, it is clear that the 
dimension of 
${\cal{S}}$ is smaller than the dimension of ${\cal{H}}$.
It is  instrumental to note  that $ - ({\mathbf{T}} + 3\, 
\bigone)$ is upto
the factor $1/\tau$ equal to the jump matrix 
${\mathbf{A}}$ for 
the translational diffusion on the square lattice 
${\cal{L}} = ([1,\ell] 
\cap {\mathbb{N}}\,)^{2}$,
which we know to diagonalize due to its translational
symmetry  when we assume cyclic boundary conditions.
For other types of correlations than given by Eq. (\ref{Eq1})
an analogous argument will hold.
 The eigenvalues of ${\mathbf{A}}$ are
$- 4 \sin^{2} \,[\,\frac{\pi}{\ell}\,(\,k_{x} - 1\,)\,] - 
4 \sin^{2}\,[\, \frac{\pi}{\ell}\,(\,k_{y} - 1\,)\,]$,
hence the eigenvalues of ${\mathbf{T}}$ are
$- 3 + 4 \sin^{2}\,[\, \frac{\pi}{\ell}\,(\,k_{x} - 1\,)\,] +
4 \sin^{2}\,[\, \frac{\pi}{\ell}\,(\,k_{y} - 1\,)\,]$,
such that:

\begin{equation} \label{Eq4}
\det{\mathbf{T}} = \prod_{k_{x}=1}^{\ell}\, \prod_{k_{y}=1}^{\ell}\,
\{\,- 3 + 4 \sin^{2}\,[\, \frac{\pi}{\ell}\,(k_{x}-1)\,] +
 4 \sin^{2}\,[\, \frac{\pi}{\ell}\,(k_{y}-1)\,] \,\}.
\end{equation}

\noindent If one of the factors that occurs in this product is zero,
we will know that the problem is non-ergodic.
E.g. for $\ell=12 \, \& \, k_{x} = 3 \, \& \, k_{y} = 4$, we have
$- 3 + 4 \sin^{2}\,[\, \frac{\pi}{\ell}\,(\,k_{x} - 1\,) \,] +
 4 \sin^{2}\,[\, \frac{\pi}{\ell}\,(\,k_{y} - 1\,) \,] = 0$,
such that in all cases where $12$ is a factor of $\ell$, 
the determinant will be zero,
leading to a reduction of the dimensionality of the problem.
We can infer from this that our dynamics very often will not
be ergodic when we adopt cyclic boundary conditions on ${\cal{L}}$
We will therefore sketch a non-rigorous argument
that suggests that for large samples non-ergodicity is not
important.

\subsection{A loophole of escape}

 Let us
imagine that the neutron beam only illuminates a 
small part (${\cal{L}}$) of the
whole  sample (${\cal{L}}'$). We have then 
${\cal{L}} \subset {\cal{L}}'$. We assume that
the dynamics on ${\cal{L}}'$ are not ergodic.
 Let us call ${\cal{H}}'$ the
configuration space that can be constructed
from ${\cal{L}}'$ and 
${\cal{H}}_{1}$ the
configuration space that could be constructed
by independent
dynamics from ${\cal{L}}$, if it were
an isolated piece of sample.
Let us drop from
the description of the configurations in ${\cal{H}}'$
everything that cannot be seen by the neutron beam,
i.e. restrict their description to ${\cal{L}}$.
Very often this incomplete, truncated description  will
generate all possible configurations of ${\cal{L}}$,
i.e. the whole of ${\cal{H}}_{1}$.
In other words, the system is then not ergodic 
but the neutron beam cannot see it:
${\cal{L}}$ has a (fake) apparent ergodicity.
This boils down to a weakening of the ergodicity
from a global to a local criterium. 
On any sizeable patch of the sample all possible
configurations will then occur.

We will be in such a case of local ergodicity if we can build
from a linear combination with integer coefficients
of vectors ${\mathbf{v}}_{\,{\mathbf{k}}}$, 
a configuration ${\mathbf{c}}_{\,{\mathbf{j}}} \in {\cal{H}}'$,
whose restriction to ${\cal{L}}$
is equivalent to  ${\mathbf{e}}_{\,{\mathbf{j}}}$, for some 
${\mathbf{j}} \in {\cal{L}}$;
put differently, if the projection of this configuration 
onto ${\cal{H}}_{1}$
is ${\mathbf{e}}_{\,{\mathbf{j}}}$. By translational 
invariance, we will then be able to
do this $\forall {\mathbf{j}} \in {\cal{L}}$, and thus 
we will also be able to
construct by linear combination any configuration 
${\mathbf{c}} \in {\cal{H}}_{1}$,
i.e. we can build a set whose projection is the 
whole of ${\cal{H}}$.

In the case of Eq. (\ref{Eq1}) it looks quite 
plausible that this would be possible:
We start with ${\mathbf{v}}_{\,{\mathbf{j}}}$. 
By using ${\mathbf{v}}_{\,{\mathbf{k}}}$
contributions from the second-neighbour shell 
we can zero the coefficients of
the ${\mathbf{e}}_{\,{\mathbf{k}}}$ corresponding to
the first-neighbour shell, and so we can proceed 
further outwards
shell after shell, upto arbitrary distances from 
${\mathbf{j}}$. For the example of the dynamics 
given by Eq. (\ref{Eq1})
we have verified in detail that this works out.
There are no conclusions to be drawn from this single case
for any other posibble model.
One will have to check case per case if
one can fall back onto such an argument of local ergodicity.

\section{Appendix 2. Defining environment-dependent dynamics}

We want to show here how one can define environment-dependent 
dynamics self-consistently.
This will allow us to explain why we are not able to treat 
such dynamics by our method.
Let ${\mathbf{c}} \in {\cal{H}}$ be a configuration,
${\mathbf{c}} = \sum_{{\mathbf{j}} \in 
{\cal{L}}}\,c_{{\mathbf{j}}} \,{\mathbf{e}}_{{\mathbf{j}}}$.
If we want to jump from ${\mathbf{c}}$ to 
${\mathbf{c}} \pm {\mathbf{e}}{{\mathbf{j}}}$
we must look at the environment of the molecule at 
${\mathbf{j}}$, i.e. we must
look at the local configuration built by all the 
molecules that belong to a cluster
${\mathbf{j}} + {\cal{G}}$ centered around 
${\mathbf{j}}$, where ${\cal{G}}$ is a set
of relative position vectors. The local configuration defined 
by these molecules
is described by the coefficients $c_{{\mathbf{j}} +
{\mathbf{g}}}, {\mathbf{g}}\in {\cal{G}}$,
i.e. by $(\,c_{{\mathbf{j}} +{\mathbf{g}}_{1}}, c_{{\mathbf{j}} + 
{\mathbf{g}}_{2}}, \dots c_{{\mathbf{j}} + {\mathbf{g}}_{\ell}}, \dots 
c_{{\mathbf{j}} + {\mathbf{g}}_{\mu}}\,) \,\in ({\mathbb{Z}}/n)^{\mu}$,
where $\mu$ is the number of molecules  inside ${\cal{G}}$.
To each such multiplet of $n^{\mu}$ vectors will correspond a
relaxation time 

\begin{equation} \label{Eqaa} 
\tau(\,c_{{\mathbf{j}} +{\mathbf{g}}_{1}}, c_{{\mathbf{j}} + 
{\mathbf{g}}_{2}}, \dots c_{{\mathbf{j}} + {\mathbf{g}}_{\ell}}, \dots 
c_{{\mathbf{j}} + {\mathbf{g}}_{\mu}}\,).
\end{equation}

\noindent Note that the quantities $c_{{\mathbf{j}} + {\mathbf{g}}_{\ell}}$
are only a small subset of the coordinates of ${\mathbf{c}}$. Some of these
 relaxation times will be
equal for symmetry reasons, but the notation remains generally valid. 
This definition holds at any site ${\mathbf{j}}$, due to the 
translational invariance
on ${\cal{L}}$. The  relaxation time Eq. (\ref{Eqaa}) goes with jumps
${\mathbf{c}} \leftrightarrow {\mathbf{c}} \pm {\mathbf{e}}_{{\mathbf{j}}}$,
i.e. it goes with $-2 \delta_{{\mathbf{c}},{\mathbf{d}}} + 
\delta_{{\mathbf{c}},{\mathbf{d}}
+ {\mathbf{e}}_{{\mathbf{j}}} } + \delta_{{\mathbf{c}},{\mathbf{d}}
- {\mathbf{e}}_{{\mathbf{j}}} }$, such that we have

\begin{equation} \label{Eqab}
M_{{\mathbf{c}},{\mathbf{d}}} = \sum_{{\mathbf{j}} \in {\cal{L}}}\,
{-2 \delta_{{\mathbf{c}},{\mathbf{d}}}
 + \delta_{{\mathbf{c}},{\mathbf{d}}
+ {\mathbf{e}}_{{\mathbf{j}}} } + \delta_{{\mathbf{c}},{\mathbf{d}}
- {\mathbf{e}}_{{\mathbf{j}}} }
 \over{\tau(\,c_{{\mathbf{j}} +
{\mathbf{g}}_{1}}, c_{{\mathbf{j}} + 
{\mathbf{g}}_{2}}, \dots c_{{\mathbf{j}} + {\mathbf{g}}_{\ell}}, \dots 
c_{{\mathbf{j}} + {\mathbf{g}}_{\mu}}}\,)}.
\end{equation}

\noindent If we further would like to introduce the possibility 
that molecules 
turn simultaneously, we will end up with

\begin{equation} \label{Eqac}
M_{{\mathbf{c}},{\mathbf{d}}} = \sum_{{\mathbf{j}} \in 
{\cal{L}}}\,\sum_{r=1}^{\rho}\,
{-2 \delta_{{\mathbf{c}},{\mathbf{d}}}
 + \delta_{{\mathbf{c}},{\mathbf{d}}
+ {\mathbf{v}}_{{\mathbf{j}},r} } + \delta_{{\mathbf{c}},{\mathbf{d}}
- {\mathbf{v}}_{{\mathbf{j}},r} } 
\over{\tau^{(r)}(\,c_{{\mathbf{j}}
 +{\mathbf{g}}_{1}}, c_{{\mathbf{j}} + 
{\mathbf{g}}_{2}}, \dots c_{{\mathbf{j}} + {\mathbf{g}}_{\ell}}, \dots 
c_{{\mathbf{j}} + {\mathbf{g}}_{\mu}}}\,)}.
\end{equation}

\noindent We can appreciate from such equations that the translational
invariance on ${\cal{H}}$ is completely lost as the relaxation times
depend on ${\mathbf{c}}$. Although (atmost) ``only'' $\rho\, n^{\mu}$ 
of them have been introduced,
it is not so that we could define a unit cell of 
$\rho\,n^{\mu}$ sites on
${\cal{H}}$ and define a ``Bravais lattice'' of 
$N/(\,\rho\, n^{\mu}\,)$ cells
in analogy with the matrix diagonalization procedure for a phonon problem.
Indeed, there is translational invariance on ${\cal{L}}$  as all clusters
${\mathbf{j}} + {\cal{G}}$ are congruent, but this does not carry over to
${\cal{H}}$, e.g. ${\mathbf{e}}_{ {\mathbf{j}} + 
{\mathbf{s}} } - {\mathbf{e}}_{{\mathbf{j}}} \neq 
{\mathbf{e}}_{ {\mathbf{j}} + 2\,{\mathbf{s}} } - 
{\mathbf{e}}_{{\mathbf{j}} + {\mathbf{s}} }$,
although $(\,{\mathbf{j}} + {\mathbf{s}}\,) - {\mathbf{j}} = 
(\,{\mathbf{j}} + 2\, {\mathbf{s}}\,) - (\,{\mathbf{j}} + 
\, {\mathbf{s}}\,)$.

The loss of translational symmetry has dramatic consequences.
We can no longer diagonalize the jump matrix by simple 
arguments of translational symmetry
and we run out of methods. There are other types of symmetry present,
but these are not high enough
to simplify the problem in any significant way.
Therefore, in this case we cannot expect to recover a behaviour 
where the elastic intensity
would vanish between the Bragg peaks, and the coherent quasielastic
intensity would not reveal the correlations present.


\begin{thebibliography}{9}
\bibitem{remark} The configuration space is not a vector space,
 as the number set used will
 most of the time be of the type ${\mathbb{Z}}/n$, 
 which is not a number field.
 In a genuine vector space we cannot have 
 $(\,\exists \lambda \neq 0\,)\,
 (\,\exists {\mathbf{v}} \neq {\mathbf{0}}\,)\, 
 \lambda{\mathbf{v}} \equiv {\mathbf{0}}$.
\end{thebibliography}
\end{document}